\documentclass[journal,10pt]{IEEEtran}
\ifCLASSINFOpdf
\else
\fi
\newtheorem{remark}{Remark}

\usepackage{graphicx}

\usepackage[cmintegrals]{newtxmath}
\begin{document}
%
\title{Deep Learning Based Joint Pilot Design and Channel Estimation for Multiuser MIMO Channels}
%
%
%


\author{Chang-Jae Chun, Jae-Mo Kang, and Il-Min Kim, \IEEEmembership{Senior~Member,~IEEE}
\thanks{C.-J. Chun, J.-M. Kang, and I.-M. Kim are with the Department of Electrical and Computer Engineering, Queen's University, Kingston, ON K7L 3N6, Canada (e-mail: changjae.chun@queensu.ca; jaemo.kang@queensu.ca; ilmin.kim@queensu.ca).}
\vspace{-7mm}
}

\maketitle

\begin{abstract}
In this paper, we propose a joint pilot design and channel estimation scheme based on the deep learning (DL) technique for multiuser multiple-input multiple output (MIMO) channels.
To this end, we construct a pilot designer using two-layer neural networks (TNNs) and a channel estimator using deep neural networks (DNNs), which are jointly trained to minimize the mean square error (MSE) of channel estimation.
To effectively reduce the interference among the multiple users, we also use the successive interference cancellation (SIC) technique in the channel estimation process.
The numerical results demonstrate that the proposed scheme considerably outperforms the state-of-the-art linear minimum mean square error (LMMSE) based channel estimation scheme.
\end{abstract}


%

\section{Introduction}
Over the past decade, the multiuser multiple-input multiple-output (MIMO) system has received considerable attention as a promising solution to improving the spectral efficiency of wireless networks \cite{Soysal09}.
In order to provide such high spectral efficiency, in the multiuser MIMO system, accurate channel estimation is crucial.
In the multiuser MIMO system, channel estimation is much more challenging for the uplink than the downlink, because inter-user interference has to be handled in the uplink.\footnote{In the downlink of the multiuser MIMO system, the channel can be readily estimated by any channel estimation methods developed for the point-to-point (single user) MIMO system, because each user independently estimates (only) its own channel, without any inter-user interference, using one common pilot broadcasted by the base station (BS).}  Specifically, in the uplink, the base station (BS) receives multiple pilot signals that are simultaneously transmitted from the multiple users, which causes inter-user interference, leading to degradation of channel estimation performance.
In the literature, the issue of channel estimation for the uplink of multiuser MIMO system has been studied in several works \cite{Soysal10}--\cite{Wang15}.

Joint optimization of pilot design and channel estimation was studied for the multiuser MIMO system \cite{Soysal10} and for the multi-relay MIMO system \cite{Kang15}.
In these works, the authors developed orthogonal pilot designing methods (i.e., the matrices representing pilot transmissions from different users are orthogonal with one another) to minimize the mean square error (MSE) of channel estimation based on the linear minimum mean square error (LMMSE) estimator.
To ensure the orthogonality, they assumed that the pilot length $L$ is larger than or equal to the total number $M$ of all antennas of all users, i.e., $L \geq M$.
In the multiuser MIMO system, however, the value of $M$ can be very large because there might be many users in a cell and each of these users has typically multiple (possibly many) antennas.
Thus, for the orthogonal pilot designing, $L$ has to be (very) large as well.
This eventually leads to a significant reduction of spectral efficiency due to decreasing of data transmission time.
Therefore, it is important to develop the channel estimation methods that are working well even for the case of $L < M$. This is the main purpose of this paper.

For the case of $L < M$, joint design of the pilot signal and channel estimator was studied for a single cell \cite{Bogale14} and for multiple cells \cite{Wang15}, but only for the case of multiuser multiple-input single-output (MISO) system (i.e., the BS has multiple antennas and each user has a single antenna).
In the case of $L < M$, it is not possible to maintain the orthogonality of pilots of multiple users.
Thus, there always exists inter-user interference in the channel estimation, which eventually leads to larger values of MSE of channel estimation.
To improve the channel estimation performance, in \cite{Bogale14} and \cite{Wang15}, some (heuristic) nonorthogonal pilot design methods were developed based on the LMMSE estimator in order to mitigate the inter-user interference.
Although it was not explicitly studied in \cite{Wang15}, it is possible to extend the technique proposed in \cite{Wang15} to the multiuser MIMO system to mitigate the inter-user interference.
However, just trying to mitigate the inter-user interference by designing some heuristic methods might not  be effective, because mitigating it does not guarantee minimizing the MSE of channel estimation.
For the best performance of channel estimation, one should try to directly minimize the MSE.
To the best of our knowledge, there has been no work for pilot design to minimize the MSE of channel estimation for multiuser MIMO channels.

Another limitation in the existing schemes for joint optimization of pilot design and channel estimation is that the channel estimation method was always assumed to be {\it linear}, most typically the LMMSE estimator.
Although such assumption is very often adopted for analytical tractability, the assumption can be a very strong one that significantly restricts the channel estimation performance.
To the best of our knowledge, there has been no work that considered the possibilities of nonlinear channel estimator.

In this paper, without assuming $L\geq M$ {\it and} without assuming linearity of the estimator, we study the problem of jointly optimizing pilot design and channel estimation in order to minimize the MSE of channel estimation.
To tackle this problem, we develop a deep learning (DL) based joint pilot design and channel estimation scheme.
Specifically, we first construct two-layer neural networks (TNNs) for pilot design and we also construct deep neural networks (DNNs) for channel estimation.
Furthermore, we use the successive interference cancellation (SIC) technique at the channel estimator to further mitigate the impact of inter-user interference.
Then all those neural networks are jointly trained to minimize the MSE of channel estimation.
The designed pilots (i.e., the multiple pilot matrices for the multiple users) turn out to be {\it nonorthogonal} and the obtained channel estimator turns out to be {\it nonlinear} as well.
Through the numerical simulations, we demonstrate that the proposed scheme considerably outperforms the state-of-the-art LMMSE based channel estimation scheme.

\textit{Notation}:
We use $\otimes$ and ${\rm vec}(\cdot)$ to denote the Kronecker product and the vectorization, respectively.
Also, we use ${\rm blkdiag} (\boldsymbol{A}_1, \cdots \boldsymbol{A}_k)$ to denote a block-diagonal matrix, of which block-diagonal matrices are given by $\boldsymbol{A}_1, \cdots \boldsymbol{A}_k$.

\section{System Model and Conventional Approach}

\subsection{Joint Optimization of Pilot Design and Channel Estimation}
We consider the uplink of a multiuser MIMO system composed of one BS and $K$ users.
The $k$th user is equipped with $\tilde{M}_k$ antennas and the BS is equipped with $N$ antennas.
For channel estimation at the BS, all users simultaneously transmit their own pilot signals represented by pilot matrices $\boldsymbol{X}_k \in \mathcal{C}^{\tilde{M}_k \times L}$, $k=1,\cdots,K$, where $L$ is the time length of each pilot signal.
The transmit power constraint at the $k$th user is given by ${\rm tr}\big(\boldsymbol{X}_k\boldsymbol{X}_k^H \big) \leq p_k$, $k=1,\cdots,K$.
At the BS, the received signal $\boldsymbol{Y}\in \mathcal{C}^{N \times L}$ is given by
                    \begin{align} \label{Y}
                    \boldsymbol{Y} = \sum_{k=1}^K \boldsymbol{H}_k \boldsymbol{X}_k + \boldsymbol{Z},
                    \end{align}
where $\boldsymbol{H}_k \in \mathcal{C}^{N \times \tilde{M}_k}$ is the channel from the $k$th user to the BS and $\boldsymbol{Z} \in \mathcal{C}^{N \times L}$ is the additive noise.

In this paper, we aim to jointly optimize the pilot signals $\big\{\boldsymbol{X}_k\big\}_{k=1}^K$ of the $K$ users and the channel estimator of the BS in order to minimize the MSE of channel estimation.
To this end, we first rewrite the received signal (\ref{Y}) as a vector form:
                    \begin{align} \label{y}
                    \boldsymbol{y}= \boldsymbol{S} \boldsymbol{g} + \boldsymbol{z},
                    \end{align}
where $\boldsymbol{y} = {\rm vec}(\boldsymbol{Y}) \in \mathcal{C}^{NL \times 1}$; $\boldsymbol{S} = \big[ \boldsymbol{X}_1^T \otimes \boldsymbol{I}_N, \cdots , \boldsymbol{X}_K^T \otimes \boldsymbol{I}_N \big] \in \mathcal{C}^{NL \times NM}$; $\boldsymbol{g} = \big[ \boldsymbol{h}_1^T,\cdots,\boldsymbol{h}_K^T \big]^T \in \mathcal{C}^{NM \times 1}$; $\boldsymbol{h}_k = {\rm vec}(\boldsymbol{H}_k) \in \mathcal{C}^{N\tilde{M}_k \times 1}$; $\boldsymbol{z} = {\rm vec}(\boldsymbol{Z}) \in \mathcal{C}^{NL \times 1}$; and $M = \sum_{k=1}^K \tilde{M}_k$.
At the BS, the channel vector $\boldsymbol{g}$ (i.e., the stacked version of the $K$ channels, $\{\boldsymbol{h}_k \}_{k=1}^K$) needs to be estimated by a channel estimator $\mathcal{F}(\cdot)$ using the received signal $\boldsymbol{y}$ and the knowledge of the pilot signals $\{\boldsymbol{X}_k\}_{k=1}^K$.
Thus, the estimated channel vector, $\hat{\boldsymbol{g}}$, can be written as $\hat{\boldsymbol{g}} = \mathcal{F} \big(\boldsymbol{y}; \boldsymbol{X}_1, \cdots, \boldsymbol{X}_K \big)$.
Then, the problem of joint optimization of pilot design and channel estimation can be written as follows:
                   \begin{align}
                    {\rm (P1)}:~ \underset{\mathcal{F}(\cdot), \{\boldsymbol{X}_k\}}{\rm min}
                    &~~ \mathbb{E}\Big[  \big|\big|\boldsymbol{g} - \hat{\boldsymbol{g}}\big|\big|^2 \Big] \nonumber\\
                    {\rm s.t.}~~&~~\hat{\boldsymbol{g}} = \mathcal{F} \big(\boldsymbol{y}; \boldsymbol{X}_1, \cdots, \boldsymbol{X}_K \big), \nonumber\\
                    &~~ {\rm tr}\big(\boldsymbol{X}_k\boldsymbol{X}_k^H \big) \leq p_k,~~ k=1,\cdots,K. \nonumber
                    \end{align}

\subsection{Conventional Approach: Heuristic Design of Pilot and Linear Estimator}
In the literature, the state-of-the-art approach to solve (P1) is to assume that the channel estimator $\mathcal{F}(\cdot)$ is linear, typically LMMSE, and to design the pilots based on heuristic methods such as in \cite{Wang15}.
In the following, we extend the LMMSE based pilot designing scheme developed in \cite{Wang15} for the multiuser MISO system to the case of the multiuser MIMO system.\footnote{To the best of our knowledge, there has been no work for joint pilot design and channel estimation for the multiuser MIMO system without assuming $L\geq M$. Thus, we extend the approach of \cite{Wang15} to the multiuser MIMO case.}

Let $\hat{\boldsymbol{g}}_{\rm LMMSE} = \bar{\boldsymbol{D}} \boldsymbol{y}$ denote the channel vector estimated by an LMMSE estimator.
The linear estimator $\bar{\boldsymbol{D}}$ to minimize the MSE of channel estimation is obtained as a function of the pilot signals: $\bar{\boldsymbol{D}} = \big[ \boldsymbol{D}_1^T, \boldsymbol{D}_2^T, \cdots, \boldsymbol{D}_K^T \big]^T$, where $\boldsymbol{D}_k = \boldsymbol{R}_{h,k} (\boldsymbol{X}_k^* \otimes \boldsymbol{I}_N) \big(\sum_{k=1}^K (\boldsymbol{X}_k^T \otimes \boldsymbol{I}_N) \boldsymbol{R}_{h,k} (\boldsymbol{X}_k^* \otimes \boldsymbol{I}_N)  + \boldsymbol{C}_z \big)^{-1}$; $\boldsymbol{R}_{h,k} = \mathbb{E}\big(\boldsymbol{h}_k \boldsymbol{h}_k^H\big)$; and $\boldsymbol{C}_z = \mathbb{E}\big(\boldsymbol{z}\boldsymbol{z}^H\big)$.
Using the linear estimator $\bar{\boldsymbol{D}}$, the MSE of channel estimation can be written as $\mathcal{E}
=\mathbb{E} \big[  ||\boldsymbol{g} - \bar{\boldsymbol{D}} \boldsymbol{y} ||^2 \big] = {\rm tr} \big\{ \big(\boldsymbol{C}_h^{-1} + \boldsymbol{S}^H \boldsymbol{C}_z^{-1} \boldsymbol{S} \big)^{-1}  \big\}$, where $\boldsymbol{C}_h = {\rm blkdiag}\big(\boldsymbol{R}_{h,1} ,\cdots, \boldsymbol{R}_{h,K}\big) $.
Note that the minimum MSE is achieved when $\boldsymbol{C}_h^{-1} + \boldsymbol{S}^H \boldsymbol{C}_z^{-1} \boldsymbol{S}$ in $\mathcal{E}$ is a diagonal matrix.
However, unless making a very restricting assumption of $L\geq M$, there is no matrix $\boldsymbol{S}$ that makes $\boldsymbol{C}_h^{-1} + \boldsymbol{S}^H \boldsymbol{C}_z^{-1} \boldsymbol{S}$ a diagonal matrix due to the singularity of the matrix $\boldsymbol{S}^H \boldsymbol{C}_z^{-1} \boldsymbol{S}$.
The state-of-the-art method to determine the matrix $\boldsymbol{S}$ is the heuristic pilot design method in \cite{Wang15}, which can mitigate the inter-user interference.

Although the LMMSE based pilot design method described above performs better than other existing schemes, it suffers from limitations both in the pilot design and in the channel estimator design.
Firstly, since the pilot is designed by the heuristic method without any optimization, the MSE of channel estimation is not minimized.
One might try to use the full search method to obtain the pilot matrices minimizing the MSE; however, the computational complexity of full searching is prohibitively high.
Secondly, the LMMSE estimator itself is not optimal for multiuser MIMO channels due to inter-user interference.\footnote{The LMMSE estimator is optimal for single user MIMO Gaussian channels, in which there exists no interference.}
In the next section, to overcome these shortcomings and to enhance the channel estimation performance, we exploit the DL technique to solve the problem (P1).

\section{DL Based Joint Pilot Design and Channel Estimation}
In this section, we propose a joint pilot design and channel estimation scheme for multiuser MIMO channels based on the DL technique.

\begin{figure*}[!t]
\centering
  \includegraphics[height=.32\textwidth]{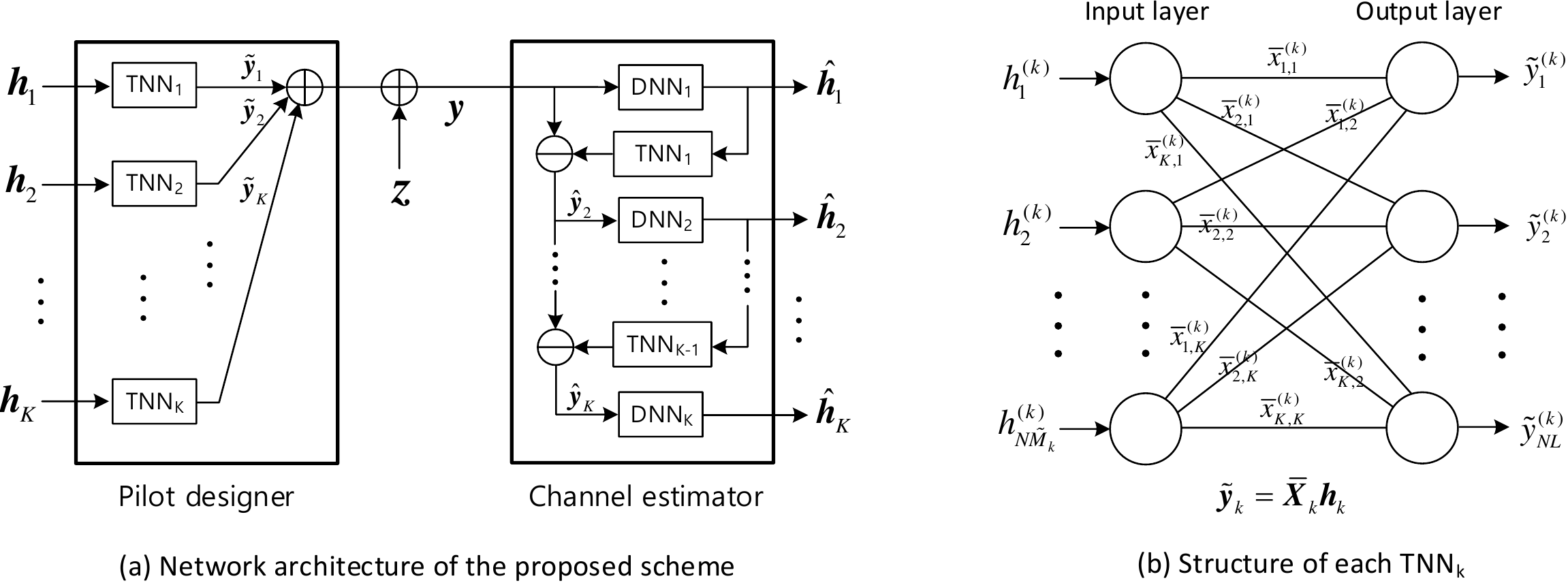}
  \caption{Structure of the proposed joint pilot design and channel estimation scheme. In Fig. \ref{Fig1}(b), $h_i^{(k)}$, $\tilde{y}_i^{(k)}$, and  $\bar{x}_{i,j}^{(k)}$ denote the $i$th element of $\boldsymbol{h}_k$, the $i$th element of $\tilde{\boldsymbol{y}}_k$, and the $(i,j)$-th element of $\bar{\boldsymbol{X}}_k$, respectively, which are the input, output, and weight of ${\rm TNN}_k$.}\label{Fig1}
  \vspace{-5mm}
\end{figure*}

\subsection{Structure and Operation of the Proposed Scheme}
The structure of the proposed scheme is shown in Fig. \ref{Fig1}(a), which is composed of two parts: the pilot designer and the channel estimator.
The goal of the pilot designer is to design $K$ pilots $\{\boldsymbol{X}_k\}_{k=1}^K$ for $K$ users to minimize the MSE by using $K$ TNNs in parallel.
The TNN used to design $\boldsymbol{X}_k$ is denoted by ${\rm TNN}_k$.
The goal of the channel estimator is to estimate $K$ channels $\{\boldsymbol{h}_k \}_{k=1}^K$ for $K$ users by iteratively stacking $K$ DNNs and $(K-1)$ TNNs in an SIC manner.
The DNN used to estimate $\boldsymbol{h}_k$ is denoted by ${\rm DNN}_k$.
The pilot designer and the channel estimator are jointly trained, as will be explained in Section III-B.
In the following, the network structure and operation of each part is first explained in detail.

\subsubsection{Pilot Designer}

In order to construct the pilot designer, letting $\bar{\boldsymbol{X}}_k = \boldsymbol{X}_k^T \otimes \boldsymbol{I}_N$, we first rewrite (\ref{y}) as follows:
                    \begin{align}\label{yK}
                    \boldsymbol{y} = \sum_{k=1}^K \tilde{\boldsymbol{y}}_k  + \boldsymbol{z},
                    \end{align}
where
                    \begin{align}\label{yt}
                    \tilde{\boldsymbol{y}}_k = \bar{\boldsymbol{X}}_k \boldsymbol{h}_k.
                    \end{align}
Interestingly, it turns out (\ref{yt}) can be modeled by a TNN as shown in Fig. \ref{Fig1}(b).
Specifically, $\boldsymbol{h}_k$ is mapped to the input to ${\rm TNN}_k$, $\bar{\boldsymbol{X}}_k$ is mapped to the weights connecting the input layer and output layer of ${\rm TNN}_k$, and $\tilde{\boldsymbol{y}}_k$ is mapped to the output of ${\rm TNN}_k$.
In all TNNs, zero bias vectors and unit activation functions are used.
Then, $\boldsymbol{y}$ is the summation of all $\tilde{\boldsymbol{y}}_k$ and the noise $\boldsymbol{z}$.
Now, one can see that (\ref{yK}), which represents the end-to-end physical mechanism of the pilot transmission and reception, can be exactly modelled by $K$ parallelly connected TNNs and the noise $\boldsymbol{z}$ as shown in Fig. \ref{Fig1}(a).

Note that $NL\times N\tilde{M}_k$ matrix $\bar{\boldsymbol{X}}_k$ is constructed by $N$ times repeatedly using all elements of $\tilde{M}_k \times L$ matrix $\boldsymbol{X}_k$, i.e., $\bar{\boldsymbol{X}}_k = \boldsymbol{X}_k^T \otimes \boldsymbol{I}_N$.
Therefore, it can be shown that the elements $\big\{\bar{x}_{i,j}^{(k)}\big\}$ of $\bar{\boldsymbol{X}}_k$ must satisfy the following two conditions:
                    \begin{align}
                    {\rm (C1)}:~&\bar{x}_{i,j}^{(k)} =  \bar{x}_{i+1,j+1}^{(k)} = \cdots = \bar{x}_{i+N-1,j+N-1}^{(k)}, \nonumber\\
                    &{\rm for}~i = 1, 1+N, \cdots, 1+(L-1)N,\nonumber\\
                    & j=1, 1+N, \cdots, 1+(\tilde{M}_k-1)N, \nonumber\\
                    {\rm (C2)}:~&\bar{x}_{i,j}^{(k)} = 0, ~~ {\rm otherwise}. \nonumber
                    \end{align}
Once training of the proposed network has completed, the pilot signals $\{\boldsymbol{X}_k\}_{k=1}^K$ in (P1) can be immediately determined by simply {\it reading} the weight matrices $\{\bar{\boldsymbol{X}}_k\}_{k=1}^K$ of the $K$ TNNs and using the relationship between $\boldsymbol{X}_k$ and $\bar{\boldsymbol{X}}_k$, i.e., $\bar{\boldsymbol{X}}_k = \boldsymbol{X}_k^T \otimes \boldsymbol{I}_N$.

\begin{remark}
In \cite{Kang18}, a single TNN was used to design a pilot for the point-to-point (single user) MISO system.
Note, however, that \cite{Kang18} cannot be used for the multiuser MIMO system, because in the multiuser MIMO system, there are multiple users and the channel for each user is given in a matrix form.
Specifically, by the approach of \cite{Kang18}, the $K$ simultaneous pilot transmissions through the $K$ channels cannot be represented, and thus, the Kronecker structure based pilot transmission and reception relationship in (\ref{yK}) cannot be modelled.
\end{remark}

\subsubsection{Channel Estimator}

To estimate the $K$ channels $\{\boldsymbol{h}_k\}_{k=1}^K$ of the $K$ users, we use $K$ DNNs motivated by the fact that the DNNs can accurately model (or approximate) complicated and nonlinear input-output mechanisms \cite{Hornik89}.
The weight matrix and the bias vector in the $v$th hidden layer (resp. the output layer) of the ${\rm DNN}_k$ are denoted by  $\boldsymbol{W}_{k,v}$ and $\boldsymbol{b}_{k,v}$ (resp. $\boldsymbol{W}_{k,o}$  and $\boldsymbol{b}_{k,o}$) for $v=1,\cdots,V_k$, respectively.
Then, the estimated channel $\hat{\boldsymbol{h}}_k$ can be mathematically written as
                    \begin{align} \label{DNN}
                    \hat{\boldsymbol{h}}_k =
                    & \boldsymbol{W}_{k,o} \phi_{k,V_k}  \Big(\boldsymbol{W}_{k,V_k} \phi_{k,V-1}\big(\cdots \phi_{k,1}(\boldsymbol{W}_{k,1}\hat{\boldsymbol{y}}_k   + \boldsymbol{b}_{k,1}) \nonumber\\
                    & \cdots \big)+\boldsymbol{b}_{k,V_k} \Big)+\boldsymbol{b}_{k,o},~k=1,\cdots,K,
                    \end{align}
where $\hat{\boldsymbol{y}}_k$ is the input vector to the ${\rm DNN}_k$ and $\phi(\cdot)_{k,v}$ denote the activation functions at the nodes of the $v$th hidden layer of the ${\rm DNN}_k$. We use the rectified linear unit (ReLU) as the activation function at the nodes of all hidden layers, i.e., $\phi_{k,v}(a) = {\rm max}(0,a)$, which has been widely used for avoiding gradient vanishing problem and computational efficiency \cite{LeCun15}.

In order to mitigate the inter-user interference, we also use the SIC technique in the channel estimation process.
The SIC technique is applied for the input signal of each DNN to reduce the amount of interference contained in the input signal.
Let $\bar{\boldsymbol{X}}_0$ and $\hat{\boldsymbol{h}}_0$ denote an $NL \times NL$ zero matrix and an $NL \times 1$ zero vector, respectively.
Then, with the SIC technique, the input signal $\hat{\boldsymbol{y}}_k$ of the ${\rm DNN}_k$ can be written as
\vspace{-2mm}
                    \begin{align} \label{y_k}
                    \hat{\boldsymbol{y}}_k = \boldsymbol{y} - \sum_{i=1}^k \bar{\boldsymbol{X}}_{i-1} \hat{\boldsymbol{h}}_{i-1},~~k=1,\cdots,K.
                    \end{align}
From (\ref{DNN}) and (\ref{y_k}), the proposed channel estimator is given as shown in Fig. \ref{Fig1}(a).

\subsection{Training of the Proposed Scheme}

We jointly train all TNNs for pilot design and all DNNs for channel estimation to minimize the MSE of channel estimation.
The loss function is set to the sample mean of squared errors as follows:
                    \begin{align} \label{loss}
                     \mathcal{J}  = \frac{1}{|\mathcal{P}|} \sum_{ \boldsymbol{g} \in \mathcal{S}}  \Big|\Big| \boldsymbol{g} -  \hat{\boldsymbol{g}}_{\rm DL} \Big|\Big|^2,
                    \end{align}
where $\mathcal{P}$ denotes the set of pilot samples; $\hat{\boldsymbol{g}}_{\rm DL}= \big[ \hat{\boldsymbol{h}}_1 ^T, \cdots, \hat{\boldsymbol{h}}_K^T\big]^T = \big[\mathcal{G}_1 (\boldsymbol{y} ; \theta_1)^T, \mathcal{G}_2(\boldsymbol{y} - \bar{\boldsymbol{X}}_1 \hat{\boldsymbol{h}}_1; \theta_2 )^T, \cdots , \mathcal{G}_K (\boldsymbol{y} - \sum_{i=1}^{K-1} \bar{\boldsymbol{X}}_i \hat{\boldsymbol{h}}_i; \theta_K )^T \big]^T$; $\mathcal{G}_k(\cdot)$ denotes the operation of the ${\rm DNN}_k$; and $\theta_k$ represents the set of the weights and biases of the ${\rm DNN}_k$.
Note that the sample mean approaches MSE, based on the law of large numbers, when the number of samples increases. The weights and biases of the DNNs are updated by the stochastic gradient descent method as
                    \begin{align}
                    \theta_k \leftarrow \theta_k - \alpha \nabla_{\theta_k} \mathcal{J},~~k=1,\cdots,K,
                    \end{align}
where $\alpha > 0 $ is the step size.
Furthermore, the weights of the TNNs must satisfy the power constraint in (P1), which can be rewritten as ${\rm tr} \big(\bar{\boldsymbol{x}}_k  \bar{\boldsymbol{x}}_k^H \big) \leq p_k$, where $\bar{\boldsymbol{x}}_k = {\rm vec}(\boldsymbol{X}_k)$.
To this end, we use the projected gradient descent method \cite{Kang18} to update the weights of the TNNs as follows:
                    \begin{align}
                    \bar{\boldsymbol{x}}_k  \leftarrow
                    \left\{
                      \begin{array}{ll}
                        \boldsymbol{u}_k,  &\textrm{if $||\boldsymbol{u}_k||^2 \leq p_k$}\\
                        \sqrt{p_k} \boldsymbol{u}_k \big/  ||\boldsymbol{u}_k||, &\textrm{if $||\boldsymbol{u}_k||^2 > p_k$}
                      \end{array}
                    \right., ~~k=1,\cdots,K,
                    \end{align}
where $\boldsymbol{u}_k = \bar{\boldsymbol{x}}_k  - \alpha \nabla_{\bar{\boldsymbol{x}}_k} \mathcal{J}$.

The proposed scheme can be trained {\it offline} using the samples of the channels and noises, which are generated according to the channel and noise statistics.
Once the proposed scheme is trained, the obtained pilot signals and the channel estimator can be effectively used for an {\it online} channel estimation process.
The proposed scheme is very useful for practical applications, because the proposed scheme can be adaptively used for various channel and noise environments by using the offline training with the appropriately sampled channels and noises.

\vspace{-3mm}
\section{Numerical Results}

\begin{figure}[!t]
\centering
  \includegraphics[height=.37\textwidth]{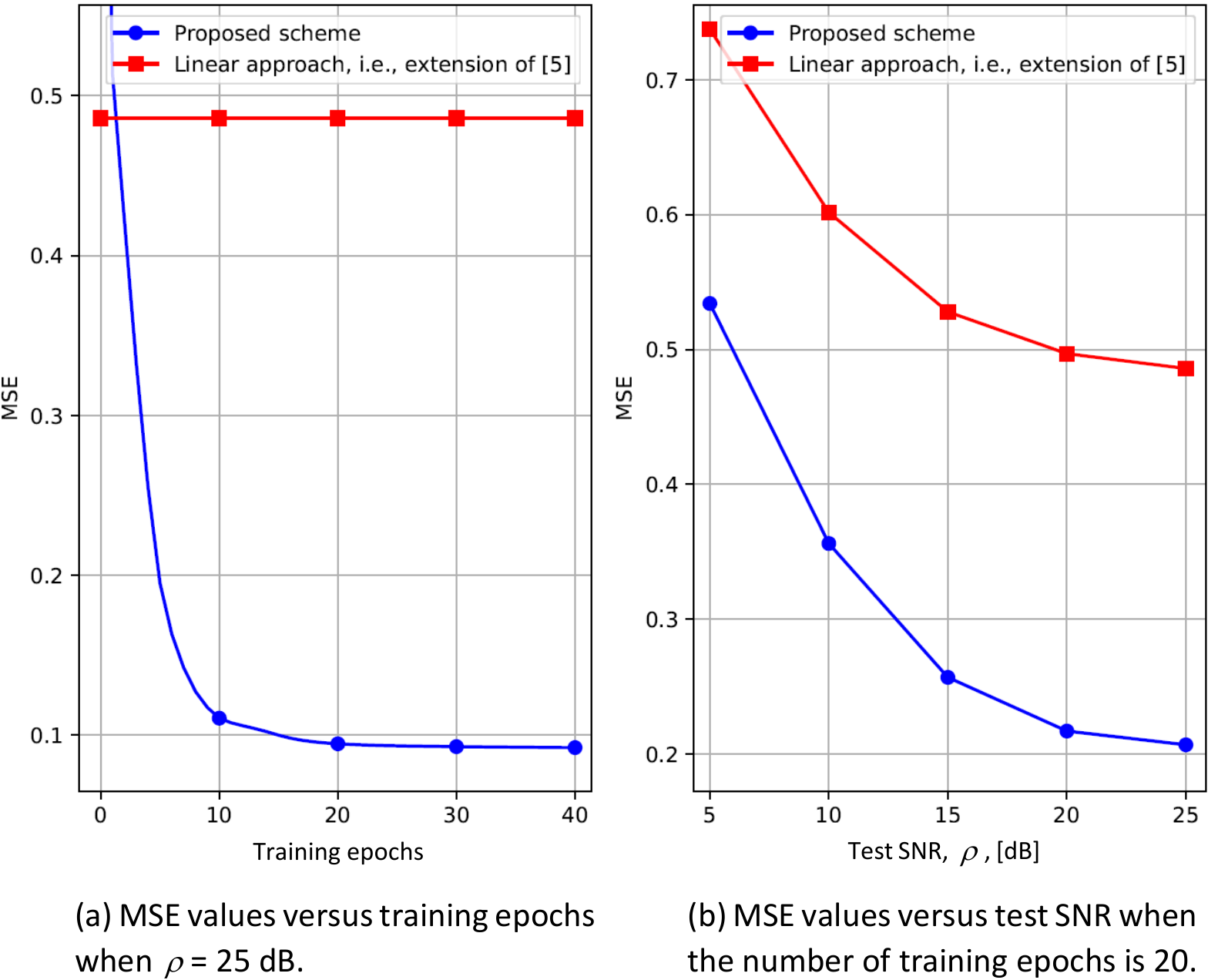}
  \caption{Channel estimation performance comparison between our proposed scheme and the state-of-the-art linear scheme.}\label{Fig2}
  \vspace{-5mm}
\end{figure}

In the simulation, we consider a multiuser MIMO channel with $K=3$, $N = 4$, $\tilde{M}_k=4$, $L=8$, and $p_k = 1$.
In the proposed scheme, we set $V_k = 5$, $\gamma = 0.001$, and the number of nodes of each hidden layer is set to $60$.
We generate the elements of channel matrix $\boldsymbol{H}_k$ and the noise vector $\boldsymbol{Z}$ according to $\mathcal{CN}(0,1)$ an $\mathcal{CN}(0,\sigma^2)$, respectively.
For the linear approach which is used for performance comparison, we set the pilot signals as
$\boldsymbol{X}_k = \sqrt{\frac{p_k}{2}}\big[\boldsymbol{U}_k,~\boldsymbol{U}_k  \big]$, where $\boldsymbol{U}_k = \big[{\rm diag} \big(1, e^{j \pi} , e^{j \frac{5\pi}{6}},e^{j \frac{2\pi}{3}}\big)\big]^{k-1}$ following the approach of \cite{Wang15}.
We train the proposed scheme using $10^6$ training samples.
One training epoch is defined as one complete cycle of training by using whole training samples.
Also, we use $10^5$ test samples apart from the training samples for performance measurement.
The signal-to-noise ratio (SNR) of the $k$th user is $\rho_k = \frac{ p_k}{\sigma^2 L}$.
Given the SNR $\rho$, the SNR of each user is set to be $\rho_1 = \rho + 3~{\rm dB}$, $\rho_2 = \rho$, and $\rho_3 = \rho-3$ dB.

In Fig. \ref{Fig2}(a), the MSE of channel estimation is shown versus training epochs for $\rho=25$ ${\rm dB}$.
It can be observed that the MSE of the proposed scheme decreases as the number of gradient steps increases, meaning that the training of the neural networks proceeds well.
Also, it can be seen that the MSE performance of the proposed scheme becomes stable approximately after 20 training epochs, which
demonstrates the convergence of the proposed channel estimation scheme.
After about $3$ epochs, the proposed scheme starts to outperform the scheme extended from \cite{Wang15}, which can be considered as the state-of-the-art linear scheme.

In Fig. \ref{Fig2}(b), the MSE of channel estimation is shown versus the test SNR $\rho$.
The MSE values of the proposed scheme are obtained by applying the test samples to the TNNs and DNNs, which are trained for $20$ epochs.
It can be seen that the proposed scheme considerably outperforms the state-of-the-art linear scheme.
This is because the inter-user interference is substantially reduced by using the proposed channel estimation scheme.
Also, the performance gap between the proposed scheme and the conventional scheme increases as the SNR increases.
This implies that, as the SNR increases, the proposed scheme better learns the channel coefficients by further reducing the inter-user interference.

\vspace{-2mm}
\section{Conclusion}
We proposed a DL based joint pilot design and channel estimation scheme for multiuser MIMO channels to minimize the MSE of channel estimation. The superior channel estimation performance was demonstrated through the numerical results.

%



\ifCLASSOPTIONcaptionsoff
  \newpage
\fi

\end{document}